\def\BibTeX{{\rm B\kern-.05em{\sc i\kern-.025em b}\kern-.08emT\kern-.1667em\lower.7ex\hbox{E}\kern-.125emX}}
\documentclass[sigconf, screen]{acmart}

\copyrightyear{2019} 
\acmYear{2019} 
\acmConference[MM '19]{Proceedings of the 27th ACM International Conference on Multimedia}{October 21--25, 2019}{Nice, France}
\acmBooktitle{Proceedings of the 27th ACM International Conference on Multimedia (MM '19), October 21--25, 2019, Nice, France}
\acmPrice{15.00}
\acmDOI{10.1145/3343031.3351014}
\acmISBN{978-1-4503-6889-6/19/10}

\usepackage{graphicx}
\usepackage{subfigure}
\usepackage{booktabs} % For formal tables
\usepackage{multirow}
\usepackage{euscript}
\usepackage{amssymb}
\usepackage{algorithm,algpseudocode}
\usepackage{enumitem}
\usepackage{colortbl}
\definecolor{mygray}{gray}{.9}
\setcopyright{acmcopyright}
\acmArticle{4}
\acmPrice{15.00}

\usepackage{dutchcal}
\begin{document}
\title[Comyco]{$\mathbcal{Comyco}$: Quality-Aware Adaptive Video Streaming via \\ Imitation Learning}

\author{Tianchi Huang$^{1,3}$, Chao Zhou$^{{2}*}$, Rui-Xiao Zhang$^{1,3}$, Chenglei Wu$^{1,3}$, Xin Yao$^{1,3}$, Lifeng Sun$^{{1,3}*}$}
\affiliation{$^{1}$Dept. of Computer Science and Technology, Tsinghua University}
\affiliation{$^{2}$Beijing Kuaishou Technology Co., Ltd., China}
\affiliation{$^{3}$BNRist, Dept. of Computer Science and Technology, Tsinghua University}

\affiliation{\{htc17@mails.,sunlf@\}tsinghua.edu.cn, zhouchao@kuaishou.com}

\renewcommand{\shortauthors}{Huang et al.}
\renewcommand{\authors}{Tianchi Huang, Chao Zhou, Rui-Xiao Zhang, Chenglei Wu, Xin Yao, Lifeng Sun}
\renewcommand{\title}{Comyco: Quality-Aware Adaptive Video Streaming via Imitation Learning}

\begin{abstract}

Learning-based Adaptive Bit Rate~(ABR) method, aiming to learn outstanding strategies without any presumptions, has become one of the research hotspots for adaptive streaming. 
However, it typically suffers from several issues, i.e., low sample efficiency and lack of awareness of the video quality information.
In this paper, we propose \emph{Comyco}, a video quality-aware ABR approach that enormously improves the learning-based methods by tackling the above issues. 
Comyco trains the policy via imitating expert trajectories given by the instant solver, which can not only avoid redundant exploration but also make better use of the collected samples.
Meanwhile, Comyco attempts to pick the chunk with higher perceptual video qualities rather than video bitrates. 
To achieve this, we construct Comyco's neural network architecture, video datasets and QoE metrics with video quality features.
Using trace-driven and real world experiments, we demonstrate significant improvements of Comyco's sample efficiency in comparison to prior work, with 1700x improvements in terms of the number of samples required and 16x improvements on training time required.
Moreover, results illustrate that Comyco outperforms previously proposed methods, with the improvements on average QoE of 7.5\% - 16.79\%. Especially, Comyco also surpasses state-of-the-art approach Pensieve by 7.37\% on average video quality under the same rebuffering time. 
%we show that Comyco outperforms the state-of-the-art learning-based ABR scheme Pensieve within only 30 minutes and 100 epochs, which is 0.17\% of the total training epochs and 6.25\% of the total training time compared to previous learning-based ABR approaches. 
%Comyco also rollouts near-optimal trajectories under various network conditions and videos by considering the strategy with the combination of video content features, network status as well as video playback states.

\end{abstract}

\keywords{Imitation Learning, Quality-aware, Adaptive Video Streaming}
\begin{CCSXML}

<ccs2012>
<concept>
<concept_id>10002951.10003227.10003251.10003255</concept_id>
<concept_desc>Information systems~Multimedia streaming</concept_desc>
<concept_significance>300</concept_significance>
</concept>
<concept>
<concept_id>10003033.10003068.10003073.10003075</concept_id>
<concept_desc>Networks~Network control algorithms</concept_desc>
<concept_significance>300</concept_significance>
</concept>
<concept_id>10010147.10010257.10010293.10010294</concept_id>
<concept_desc>Computing methodologies~Neural networks</concept_desc>
<concept_significance>300</concept_significance>
</concept>
</ccs2012>
\end{CCSXML}

\ccsdesc[300]{Information systems~Multimedia streaming}
%\ccsdesc[300]{Networks~Network control algorithms}
\ccsdesc[300]{Computing methodologies~Neural networks}
%\ccsdesc[500]{Networks ~ Network measurement}

%\printccsdesc

\maketitle

\section{Introduction}
Recent years have seen a tremendous increase in the requirements of watching online videos~\cite{cisco}. 
Adaptive bitrate~(ABR) streaming, the method that dynamically switches download chunk bitrates for restraining rebuffering event as well as obtaining higher video qualities, has become the popular scheme to deliver videos with high quality of experience~(QoE) to the users~\cite{bentaleb2018survey}. 
Recent model-based ABR approaches~(\S\ref{sec:related}) pick the next chunk's video bitrate via only current network status~\cite{jiang2014improving}, or buffer occupancy~\cite{spiteri2016bola}, or joint consideration of both two factors\cite{yin2015control}. 
However, such heuristic methods are usually set up with presumptions, that fail to work well under unexpected network conditions~\cite{mao2017neural}. 
Thus, learning-based ABR methods adopt reinforcement learning~(RL) method to \emph{learn} the strategies without any presumptions, which outperform traditional model-based approaches. 

Nevertheless, learning-based ABR methods suffer from two key issues.
%There are currently two major problems with the learning-based ABR scheme.
While recent work~\cite{mao2017neural,DDASH} often adopts RL methods to train the neural network, such methods lack the efficiency of both collected and exploited expert samples, which leads to the inefficient training~\cite{mendonca2019guided}. 
Besides, the majority of existing ABR approaches~\cite{yin2015control, mao2017neural,akhtar2018oboe} neglect the video quality information, while perceptual video quality is a non-trivial feature for evaluating QoE~(\S\ref{sec:QoE},\cite{huang2018qarc}). Thus, despite their abilities to achieve higher QoE objectives, such schemes may generate a strategy that diverges from the actual demand~(\S\ref{sec:challenges}). 
% As a result, quality-aware learning-based ABR algorithm is still challenging to because there are currently many factors missing, including viable neural network models, feasible and high-efficiency training methodologies, dedicated video datasets based on video quality metrics, as well as quality-based QoE methods.

In this paper, 
we propose \emph{Comyco}, a novel video quality-aware learning-based ABR system, aiming to remarkably improve the overall performance of ABR algorithms via tackling the above challenges. 
Unlike previous RL-based schemes~\cite{mao2017neural}, Comyco leverages imitation learning~\cite{osa2018algorithmic} to train the neural network~(NN). 
That is because the near-optimal policy can be precisely and instantly estimated via the current state in the ABR scenario and the collected expert policies can enable the NN for fast learning. 
Following this thought~(\S\ref{sec:confrontation}), the agent is allowed to \emph{explore} the environment and \emph{learn} the policy via the expert policies given by the solver~(\S\ref{sec:trainingmethod}). 
%we can compute the best direction of gradients to update the NN faster.
Specifically, we propose \emph{instant solver}~(\S\ref{sec:instantsolver}) to estimate the expert action with a faithful \emph{virtual player}~(\S\ref{sec:abrbaseline}). Furthermore, we utilize \emph{experience replay buffer}~(\S\ref{sec:experiencereplay}) to store expert policies and train the NN via the specific loss function \emph{$L_{comyco}$}~(\S\ref{ses:loss}). 
% Thus, unlike previous RL-based ABR methods, we are able to compute the best direction of gradients to update the NN faster. 
% In details, the agent first uses the NN-based sampling module to select future bitrate via the given ABR's state. The solver then computes the near-optimal bitrate for the next chunk according to the current state. Next, we update the NN via minimizing the gap between near-optimal bitrate and selected bitrate. Finally, the agent further \emph{goes} to the next state w.r.t the picked bitrate. The method is effective and straightforward, which significantly accelerates training process and improves the performance.

Besides, Comyco aims to select bitrate with high perceptual video quality rather than high video bitrate. 
%It's quite challenging because we have neither metrics nor datasets. 
To achieve this, we first integrate the information of video contents, network status, and video playback states into the Comyco's NN for bitrate selection~(\S\ref{sec:arch}). Next, we consider using VMAF~\cite{rassool2017vmaf}, an objective full-reference perceptual video quality metric, to measure the video quality. Concurrently, we also propose a linear combination of video quality-based QoE metric that achieves the state-of-art performance on Waterloo Streaming SQoE-III~\cite{duanmu2018sqoe} dataset~(\S\ref{sec:QoE}). Finally, we collect a DASH-video dataset with various types of videos, including movies, sports, TV-shows, games, news, and music videos~(MV)~(\S\ref{sec:VideoDatasets}). 
% we have observed that the video quality of entire sessions is significantly correlated with its QoE assessments, which thereby affects the optimal strategy of ABR to enable it to provide the video chunks with maximized average video qualities rather than average video bitrates. Hence, we consider to restructure the NN architecture with video content features. In order to pursue the diversity of video content, we collect a DASH-video dataset with various types of videos including movies, sports, tv-shows, games, news, and music videos~(MV). 
%At the same time, we leverage VMAF tool to calculate the perceptual video quality of each video chunk in our dataset. 

Using trace-driven emulation~(\S\ref{sec:testbed}), we find that Comyco significantly accelerates the training process, with 1700x improvements in terms of number of samples required compared to recent work~(\S\ref{sec:pensieveretrain}). 
Comparing Comyco with existing schemes under various network conditions~(\S\ref{sec:NetworkDatasets}) and videos~(\S\ref{sec:VideoDatasets}), we show that Comyco outperforms previously proposed methods, with the improvements on average QoE of 7.5\% - 16.79\%. 
In particular, Comyco performs better than state-of-the-art learning-based approach Pensieve, with the improvements on the average video quality of 7.37\% under the same rebuffering time. 
Further, we present results which highlight Comyco's performance with different hyperparameters and settings~(\S\ref{sec:cmcstudy}).
Finally, we validate Comyco in real world network scenarios~(\S\ref{real-world}). Extensive results indicate the superiority of Comyco over existing state-of-the-art approaches.

In general, we summarize the contributions as follows:
\begin{enumerate}[leftmargin=*]
    \item[$\triangleright$] We propose Comyco, a video quality-aware learning-based ABR system, that significantly ameliorates the weakness of the learning-based ABR schemes from two perspectives.
    
    \item[$\triangleright$] To the best of our knowledge, we are the first to leverage imitation learning to accelerate the training process for ABR tasks. Results indicate that utilizing imitation learning can not only achieve fast convergence rates but also improve performance.
    
    \item[$\triangleright$] Unlike prior work, Comyco picks the video chunk with high perceptual video quality instead of high video bitrate. Results also demonstrate the superiority of the proposed algorithm.
\end{enumerate}
\section{Background and Challenges}
%In this section, we begin with introducing ABR's background. Then we propose the key challenges of learning-based ABR algorithms.

\subsection{ABR Overview}
Due to the rapid development of network services, watching video online has become a common trend. Today, the predominant form for video delivery is adaptive video streaming, such as HLS~(HTTP Live Streaming)~\cite{HLS} and DASH~\cite{dash}, which is a method that dynamically selects video bitrates according to network conditions and clients' buffer occupancy. Traditional video streaming framework consists of a video player client with a constrained buffer length and an HTTP-Server or Content Delivery Network~(CDN). The video player client decodes and renders video frames from the playback buffer. Once the streaming service starts, the client fetches the video chunk from the HTTP Server or CDN in order by an ABR algorithm. Meanwhile, the algorithm, deployed on the client side, determines the next chunk $N$ and next chunk video quality $Q_N$ via throughput estimation and current buffer utilization. The goal of the ABR algorithm is to provide the video chunk with high qualities and avoid stalling or rebuffering~\cite{bentaleb2018survey}.
%, 2)~the delay $T$ that represents when to download the next chunk $N$. 
% After finished to play the video, several metrics, such as total bitrate $b$, total re-buffering time $r$ and total bitrate change $s$ will be summarized as a QoE metric to evaluate the performance. Thus, achieving a high QoE score for video streaming has become a major challenge for ABR algorithms. 

\subsection{Challenges for learning-based ABRs}
%Meanwhile, learning-based ABR methods still remains several upcoming challenges to tackle it.
\label{sec:challenges}
Most traditional ABR algorithms~\cite{jiang2014improving,yin2015control,spiteri2016bola} leverage time-series prediction or automation control method to make decisions for the next chunk. Nevertheless, such methods are built in pre-assumptions that it is hard to keep its performance in all considered network scenarios~\cite{mao2017neural}. To this end, learning-based ABR algorithms~\cite{mao2017neural,DDASH,huang2018tiyuntsong} are proposed to solve the problem from another perspective: it adopts deep reinforcement learning~(DRL) to train a neural network~(NN) from scratch towards the better QoE objective. Despite the outstanding results that recent work has obtained, learning-based ABR methods suffer from several key issues:
%including its training methodologies and incomplete feature representations.

\noindent \textbf{The weaknesses of RL-based ABR algorithms.}~Recent learning-based ABR schemes often adopt RL methods to maximize the average QoE objectives. During the training, the agent rollouts a trajectory and updates the NN with policy gradients. However, the effect of calculated gradients heavily depends on the amount and quality of collected experiences. In most cases, the collected samples seldom stand for the optimal policy of the corresponding states, which leads to a long time to converge to the sub-optimal policy~\cite{osa2018algorithmic, mao2019variance}. Thus, we are facing the first challenge:~\emph{Considering the characteristic of ABR tasks, can we precisely estimate the optimal direction of gradients to guide the model for better updating?}

\begin{figure}
    \centering
    \begin{minipage}{0.90\linewidth}
        \centering
        \includegraphics[width=0.8\textwidth]{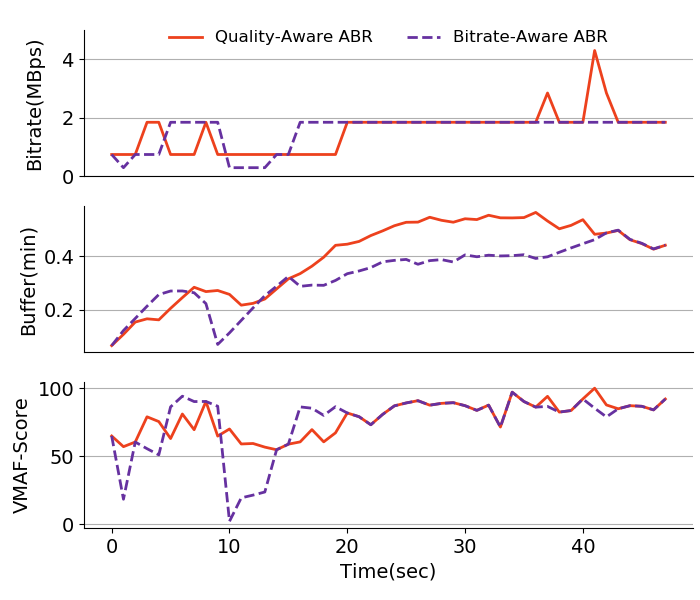}
    \end{minipage}  
    % \begin{minipage}{0.40\linewidth}
    %     \centering
    %     \subfigure[Trace No.\textbf{115} in Oboe Dataset]{\includegraphics[width=0.8\textwidth]{figs/optimal/115}}
    % \end{minipage}
    \caption{We evaluate quality-aware ABR algorithm and bitrate-aware ABR algorithm with the same video on Norway network traces respectively. Results are plotted as the curves of selected bitrate, buffer occupancy and the selected chunk's VMAF~(\S\ref{sec:QoE},\cite{rassool2017vmaf}) for entire sessions.}
    \label{fig:optimalvmafqoe}
    \vspace{-15pt}
\end{figure}

\noindent \textbf{The unique video quality.}~What's more, previous learning-based ABR schemes~\cite{yin2015control, mao2017neural} are evaluated by typical QoE objectives that use the combination of video bitrates, rebuffering times and video smoothness. However, such QoE metrics are short-handed because these forms of parameters neglect the quality of video presentations~\cite{zhou_wang_2017}. Meanwhile, recent work~\cite{qin2018abr, duanmu2017quality} has found that perceptual video quality features play a vital part in evaluating the performance of VBR-encoded ABR streaming services. To prove this, we plot the trajectory generated by the quality-aware ABR and bitrate-aware algorithm on Figure~\ref{fig:optimalvmafqoe}. As shown, the bitrate-aware algorithm selects the video chunk with higher bitrate but neglects the corresponding video quality, resulting in a large fluctuation in the perceptual video qualities. What's more, bitrate-aware algorithm often wastes the buffer on achieving a slight increase in video quality, which may cause unnecessary stalling event. On the contrast, the quality-aware algorithm picks the chunk with high and stable perceptual video quality and preserves the buffer occupancy within an allowable range.
% typical QoE models leverages the video bitrate, rebuffering times and frequency as the key parameters. It's obvious that such QoE metrics are short-handed because these forms of parameters neglects the presentation quality of the video~\cite{duanmu2017quality}. 
To this end, one of the better solutions is to add video bitrates as another metric to describe the perceptual video quality. We, therefore, encounter the second challenge of our work:~\emph{How to construct a video quality-aware ABR system?}

% Hence, in this paper, our goal is to 

\section{Methods}
Motivated by the key challenges~(\S\ref{sec:challenges}), we propose Comyco, a video quality-aware learning-based ABR scheme. In this section, we introduce two main ideas of Comyco: training NN via imitation learning~(\S\ref{sec:confrontation}) and a complete video quality-based ABR system~(\S\ref{sec:qualityaware}).
\begin{figure}
    \centering
    \begin{minipage}{0.45\linewidth}
        \centering
        \subfigure[Supervised learning]{\includegraphics[width=1.0\textwidth]{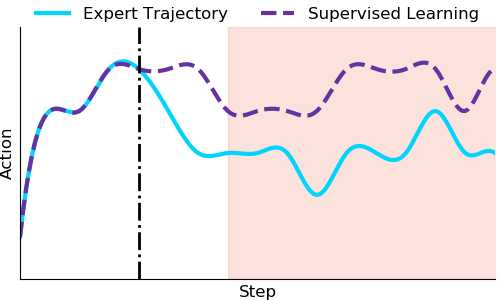}}
    \end{minipage}  
    \begin{minipage}{0.45\linewidth}
        \centering
        \subfigure[Imitation learning]{\includegraphics[width=1.0\textwidth]{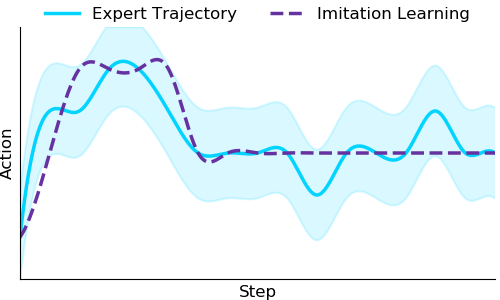}}
    \end{minipage}
    \caption{The real trajectory on the ABR task given by imitation learning and supervised learning, where the red background means the player occurs the rebuffering event.} 
    %supervised-learning aims to \emph{copy} }
    \label{fig:imitation}
    %\vspace{-20pt}
\end{figure}

\subsection{Training ABRs via Imitation Learning}
% The key principle of RL-based method is to maximize \emph{reward} of each \emph{action} taken by the agent in given \emph{states} per step. Recall that RL-based methods are talented in the circumstance that is difficult to infer the best action for the given state, such as Go, StarCraft II, and Atari games. While 
% RL-based methods
\label{sec:confrontation}
Recall that the key principle of RL-based method is to maximize \emph{reward} of each \emph{action} taken by the agent in given \emph{states} per step, since the agent doesn't really know the optimal strategy~\cite{sutton2018reinforcement}. However, recent work~\cite{yin2015control,mao2017neural,spiteri2018theory,akhtar2018oboe,pereira2018cross,huang2018tiyuntsong} has demonstrated that the ABR process can be precisely emulated by an offline virtual player~(\S\ref{sec:testbed}) with complete future network information. What's more, by taking several steps ahead, we can further accurately \emph{estimate} the near-optimal expert policy of any ABR state within an acceptable time~(\S\ref{sec:instantsolver}). 
To this end, 
%we attempt to collect the optimal policy ABR dataset and train the model via supervised learning methods. 
the intuitive idea is to leverage supervised learning methods to minimize the loss between the predicted and the expert policy.
Nevertheless, it's impractical because the off-policy method~\cite{sutton2018reinforcement} suffers from \emph{compounding error} when the algorithm executes its policy, leading it to drift to new and unexpected states~\cite{laskey2017dart}. For example, as shown in Figure~\ref{fig:imitation}[a], in the beginning, supervised learning-based ABR algorithm fetches the bitrate that is consistent with the expert policy, but when it selects a bitrate with a minor error~(after the black line), the state may be transitted to the situation not included in the dataset, so the algorithm would select another wrong bitrate. Such compounding errors eventually lead to a continuous rebuffering event. As a result, supervised-learning methods cannot learn to recover from failures. 

In this paper, we aim to leverage imitation learning, a method that closely related to RL and supervised learning, to learn the strategy from the expert policy samples. Imitation learning method reproduces desired behavior according to expert demonstrations~\cite{osa2018algorithmic}. The key idea of imitation learning is to allow the NN to explore environments and collect samples~(just like RL) and learn the policy based on the expert policy~(just as supervised learning). In detail, at step $t$, the algorithm infers a policy $\pi_t$ at ABR state $S_t$. It then computes a loss $l_t(\pi_t, \pi^{*}_t)$ w.r.t the expert policy $\pi^{*}_t$. After observing the next state $S_{t+1}$, the algorithm further provides a different policy $\pi_{t+1}$ for the next step ${t+1}$ that will incur another loss $l_t(\pi_{t+1}, \pi^{*}_{t+1})$. Thus, for each $\pi_t$ in the class of policies $T \in \{\pi_0,\dots,\pi_t\}$, we can find the policy $\hat{\pi}$ through any supervised learning algorithms~(Eq.~\ref{eq:pi}).
%~($\hat{\pi} = \mathop{arg\,min}\limits_{\pi \in T}\mathbb{E}_{s\sim d_{\pi}}\left[l_t(\pi_t, \pi^{*}_t)\right]$). 

\begin{small}
\begin{equation}
    \label{eq:pi}
    \hat{\pi} = \mathop{arg\,min}\limits_{\pi \in T}\mathbb{E}_{s\sim d_{\pi}}\left[l_t(\pi_t, \pi^{*}_t)\right]
\end{equation}
\end{small}

Figure~\ref{fig:imitation}[b] elaborates the principle of imitation learning-based ABR schemes: the algorithm attempts to explore the strategy in a range near the expert trajectory to avoid compounding errors.
%To this end, as shown in Figure~\ref{fig:imitation}[b], 

\subsection{Video Quality-aware ABR System Setup}
\label{sec:qualityaware}
Our next challenge is to set up a video quality-aware ABR system. The work is generally composed of three tasks: 1)~We construct Comyco's NN architecture with jointly considering several underlying metrics, i.e, past network features and video content features as well as video playback features~(\S\ref{sec:arch}). 2)~We propose a quality-based QoE metric~(\S\ref{sec:QoE}). 3)~We collect a video quality DASH dataset which includes various types of videos~(\S\ref{sec:VideoDatasets}).
\section{System Overview}
\begin{figure}
    \centering
    \includegraphics[width=0.9\linewidth]{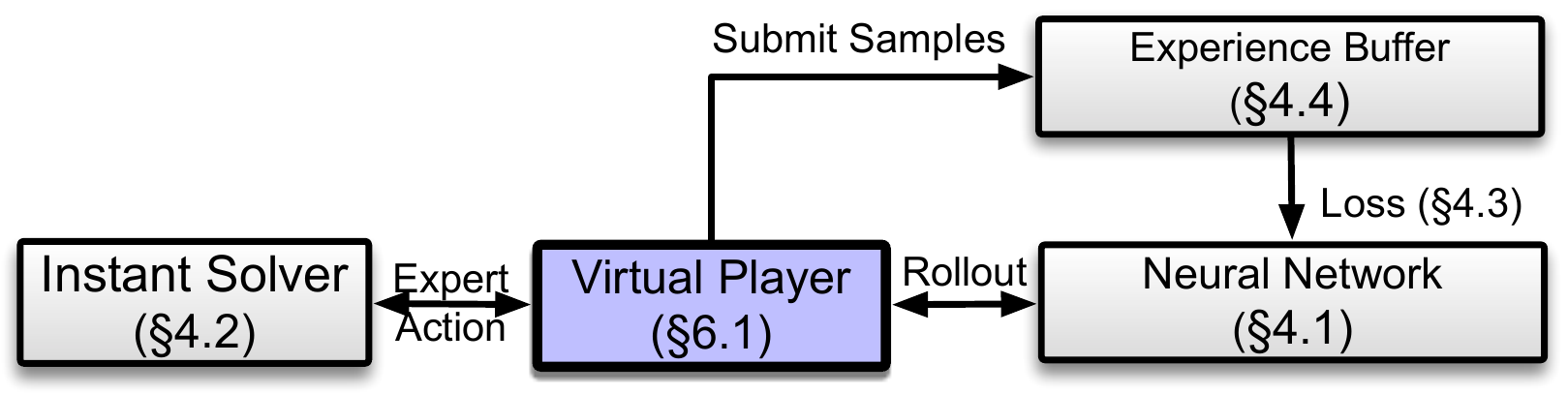}
    \caption{Comyco's Basic System Work-flow Overview. Training methodologies are available in \S\ref{sec:trainingmethod}.}
    \label{fig:overview}
  %   \vspace{-10pt}
\end{figure}

In this section, we describe the proposed system in detail. Comyco's basic system work-flow is illustrated in Figure~\ref{fig:overview}. The system is mainly composed of a NN, an ABR virtual player, an instant solver, and an experience replay buffer. We start by introducing the Comyco's modules. Then we explain the basic training methodology. Finally, we further illustrate Comyco with a multi-agent framework.

\subsection{NN Architecture Overview}
\label{sec:arch}
Motivated by the recent success of on-policy RL-based methods, Comyco's learning agent is allowed to explore the environment via traditional rollout methods. For each epoch $t$, the agent aims to select next bitrate via a neural network~(NN). We now explain the details of the agent's NN including its inputs, outputs, network architecture, and implementation.

\noindent \textbf{Inputs.}
We categorize the NN into three parts, network features, video content features and video playback features~($S_k=\{C_k, M_k, F_k\}$). Details are described as follows.
\begin{itemize}[leftmargin=*]
    \item[$\triangleright$] \textbf{\emph{Past Network features.}}~The agent takes past $t$ chunks' network status vector $C_k=\{c_{k-t-1}, \dots, c_k\}$ into NN, where $c_i$ represents the throughput measured for video chunk $i$. Specifically, $c_i$ is computed by $c_i=n_{r,i}/{d_i}$, in which $n_{r,i}$ is the downloaded video size of chunk $i$ with selected bitrates $r$, and $d_i$ means download time for video chunk $n_{r,i}$.
    \item[$\triangleright$] \textbf{\emph{Video content features.}}~Besides that, we also consider adding video content features into NN's inputs for improving its abilities on detecting the diversity of video contents. In details, the learning agent leverages $M_k=\{N_{k+1}, V_{k+1}\}$ to represent video content features. Here $N_{k+1}$ is a vector that reflects the video size for each bitrate of the next chunk $k+1$, and $V_{k+1}$ is a vector which stands for the perceptual video quality metrics for each bitrate of the next chunk. 
    \item[$\triangleright$] \textbf{\emph{Video playback features.}}~The last essential feature for describing the ABR's state is the current video playback status. The status is represented as $F_k = \{v_{k-1}, B_k, D_k, m_k\}$, where $v_{k-1}$ is the perceptual video quality metric for the past video chunk selected, $B_k, D_k$ are vectors which stand for past t chunks' buffer occupancy and download time, and $m_k$ means the normalized video chunk remaining.
\end{itemize}

\noindent \textbf{Outputs.}
Same as previous work, we consider using discrete action space to describe the output. Note that the output is an n-dim vector indicating the probability of the bitrate being selected under the current ABR state $S_k$.

\begin{figure}
    \centering
    \includegraphics[width=0.9\linewidth]{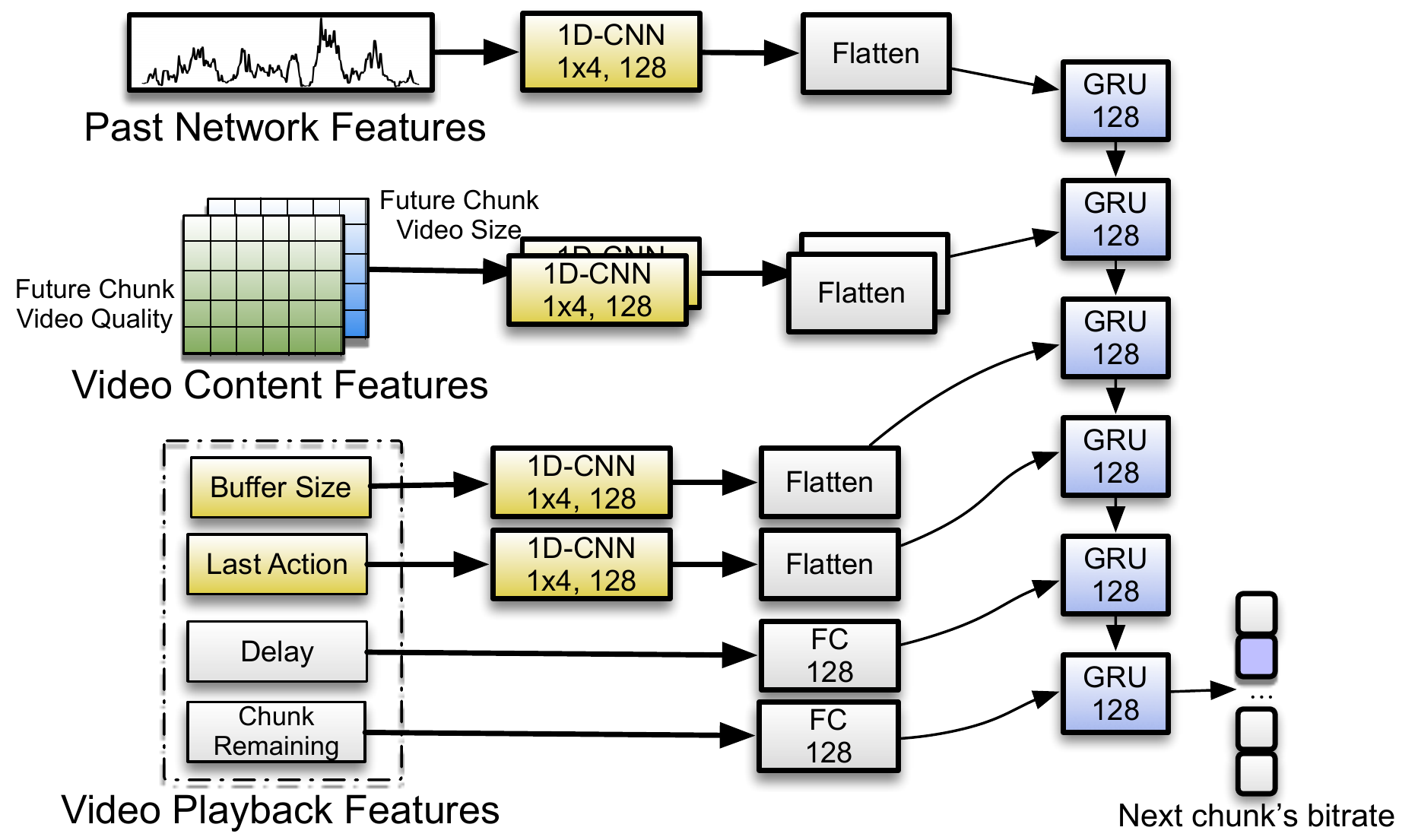}
    \caption{Comyco's NN architecture Overview.}
    \label{fig:comycoarch}
  %   \vspace{-10pt}
\end{figure}

\noindent \textbf{Implementation.}
As shown in Figure~~\ref{fig:comycoarch}, for each input type, we use a proper and specific method to extract the underlying features. In details, we first leverage a single 1D-CNN layer with kernel=4, channels=128, stride=1 to extract network features to a 128-dim layer. We then use two 1D-CNN layers with kernel=1x4, channels=128 to fetch the hidden features from the future chunk's video content matrix. Meanwhile, we utilize 1D-CNN or fully connected layer to extract the useful characteristics from each metric upon the video playback inputs. The selected features are passed into a GRU layer and outputs as a 128-dims vector. Finally, the output of the NN is a \emph{6-dims} vector, which represents the probabilities for each bitrate selected. We utilize \emph{RelU} as the active function for each feature extraction layer and leverage \emph{softmax} for the last layer.

\subsection{Instant Solver}
\label{sec:instantsolver}
Once the sampling module rolls out an action $a_t$, we aim to design an algorithm to fetch all the \emph{optimal} actions $\hat{a_t}$ with respect to current state $s_t$. Followed by these thoughts, we further propose \emph{Instant Solver}. The key idea is to choose future chunk $k$'s bitrate $R_k$ by taking $N$ steps ahead via an offline \emph{virtual player}, and solves a specific \emph{QoE maximization problem} with future network throughput measured $C_t$, in which the future real throughput can be successfully collected under both offline environments and real-world network scenarios. Inspired by recent model-based ABR work~\cite{yin2015control}, we formulate the problem as demonstrated in Eq.~\ref{eq:mpc}, denoted as $QoE^{N}_{max} K$. In detail, the virtual player consists of a virtual time, a real-world network trace and a video description. At virtual time $t_k$, we first calculate download time for chunk $k$ via $d_k(R_k) / C_k$, where $d_k$ is the video chunk size for bitrate $R_k$, and $C_k$ is average throughput measured. We then update $B_{k+1}$ buffer occupancy for chunk $k+1$, in which $\delta t_k$ reflects the waiting time such as Round-Trip-Time~(RTT) and video render time, and $B_{max}$ is the max buffer size. Finally, we refresh the virtual time $t_{k+1}$ for the next computation. Note that the problem can be solved with any optimization algorithms, such as memoization, dynamic programming as well as Hindsight~\cite{Huang2019Hindsight}. Ideally, there exists a trade-off between the computation overhead and the performance. We list the performance comparison of instant solver with different $N$ in \S\ref{sec:cmcstudy}. In this work, we set $N=8$.
\begin{small}
\begin{align}
\label{eq:mpc}
&\max \limits_{R_1,\dots,R_k,T_s}{QoE}^{N}, \qquad s.t. \notag \\
&\left\{
\begin{aligned}
&t_{k+1} = t_{k} + \frac{d_k(R_k)}{C_k} + \delta t_k, \\
&C_k =\frac{1}{t_{k+1}-t_{k}-\delta t_{k}} \int^{t_{k+1} - \delta t_k}_{t_k} C_t dt, \\
&B_{k+1} = \left[ \left( B_{k} -\frac{d_k(R_k)}{C_k}\right)_{+} + L - \delta t_k \right] _{+}, \\
&B_{1} = T_{s}, \\
&B_k \in [0, B_{max}], R_k \in R, \forall k=1, \dots, N.
\end{aligned}
\right.
\end{align}
\end{small}
\subsection{Choice of Loss Functions for Comyco}
\label{ses:loss}
%Until now we have not discussed how to define update the model. 
In this section, we start with designing the loss function from the fundamental RL training methodologies. The goal of the RL-based method is to maximize the Bellman Equation, which is equivalent to maximize the value function~$q_{\pi}(s,a)$~\cite{sutton2018reinforcement}. The equation is listed in~Eq.~\ref{eq:rl}, where $q_{*}(s,a)$ stands for the maximum action value function on all policies, $V_\pi(s)$ is the value function, $\pi(s,a;\theta)$ is the rollout policy. Thus, given an expert action $q_{\pi}(s,\hat{a})=q_{*}(s,a)$, we can update the model via minimizing the gap between the true action probability $\hat{A}$ and $\pi$, where $A$ is an one hot encoding in terms of $\hat{a}$. In this paper, we use cross entropy error as the loss function. Recall that the function can be represented as any traditional behavioral cloning loss methods~\cite{osa2018algorithmic}, such as Quadratic, LI-loss and Hinge loss function. In addition, we find that the other goal of the loss function is to maximize the probabilities of the selected action, while the goal significantly reduces the aggressiveness of exploration, and finally, resulting in obtaining the sub-optimal performance. Thus, motivated by the recent work on RL~\cite{mnih2016asynchronous}, we add the entropy of the policy $\pi$ to the loss function. It can encourage the algorithm to increase the exploration rate in the early stage and discourage it in the later stage. The loss function for Comyco is described in Eq~\ref{eq:cel}.
%Traditional loss function for behavioral cloning is summarized into four types: 
\begin{small}
\begin{equation}
    \label{eq:rl}
    \begin{aligned}
    \max \limits_{\pi}V_\pi(s) &=\sum_{a \in A}\pi(a|s;\theta)q_{\pi}(s,a) \\
    &=\max \limits_{\pi} \max \limits_{a} q_{\pi}(s,a) \\
    &=\max \limits_{a} q_{*}(s,a)
    \end{aligned}
\end{equation}

\begin{equation}
    \label{eq:cel}
    L_{comyco}=-\sum \hat{A} \log \pi(s,a;\theta) - \alpha H(\pi(s;\theta)).
\end{equation}
\end{small}
Here $\pi(s,a;\theta)$ is the rollout policy selected by the NN, $\hat{A}$ is the real action probability vector generated by the expert actor $\hat{a}$, $H(\pi(s;\theta)$ represents the entropy of the policy, $\alpha$ is a hyperparameter which controls the encouragement of exploration. In this paper, we set $alpha=0.001$ and discuss $L_{comyco}$ with different $\alpha$ in \S\ref{sec:cmcstudy}.

\subsection{Training Comyco with Experience Replay}
\label{sec:experiencereplay}
%Motivated by the recent success of off-policy~\cite{sutton1998reinforcement} 
Recent off-policy RL-based methods~\cite{mnih2013playing} leverage experience replay buffer to achieve better convergence behavior when training a function approximator. Inspired by the success of these approaches, we also create a sample buffer which can store the past expert strategies and allow the algorithm to randomly picks the sample from the buffer during the training process. We will discuss the effect of utilizing experience replay on Comyco in \S\ref{sec:cmcstudy}.

\subsection{Methodology}
\label{sec:trainingmethod}
% \noindent\textbf{Useless operation definition.}~Recall that our goal is to train an NN model for achieving higher QoE performance with less buffer-bound changes. It's quite apparent that the client will gain most QoE score if it continuously requests the new buffer-bound for replacing the old one. Nevertheless, the action is the bitrates selected from the newer buffer-bound is equivalent to the one selected from the older buffer-bound, we define this form of action as \emph{useless operation}. 
% Going along with the following idea, we create a large corpus of datasets which composed of Trigger's inputs and its ground truth for training and validating. Unfortunately, without expected, the model cannot be trained via supervised learning.

We summarize the Comyco's training methodology in Alg.~\ref{alg:TriggerOverall}. 
% As shown, Comyco's workflow mainly consists of four parts: 
% \begin{enumerate}
%     \item Samples an action via NN for the state observed.
%     \item Computes the expert action via instant solver according to the current state.
%     \item Updates the gradient by using loss function~(Eq.~\ref{eq:cel}).
%     \item Uses virtual player to emulate the next state.
% \end{enumerate}

%Specifically, the training step will be looped until the Trigger converges, or, the training process is somewhat endless.

\begin{algorithm}
\caption{Overall Training Procedure} 
\label{alg:TriggerOverall} 
\begin{small}
\begin{algorithmic}[1]
\Require Training model $\pi$, \texttt{Instant Solver}(\S\ref{sec:instantsolver}).
\State Sample Training Batch $B = \{\}$.
\Procedure{Training}{}
\State Initialize $\pi$.
\State Get State ABR state $s_t$.
\Repeat
\State Picks $a_t$ according to policy $\pi(s_t; \theta)$.
\State Expert action $\hat{a_t} = \texttt{Instant Solver} (s_t)$.
\State $B \gets B \bigcup \{s_t,\hat{a_t}\}$.
\State Samples a batch $\hat{B} \in B$.
\State Updates network $\pi$ with $\hat{B}$ using Eq.\ref{eq:cel};
\State Produces next ABR state $S_{t+1}$ according to $s_t$ and $a_t$.
\State $t \gets t+1$
\Until{Converged}
\EndProcedure

\end{algorithmic}
\end{small}
\end{algorithm}

\begin{figure}
    \centering
    \includegraphics[width=0.6\linewidth]{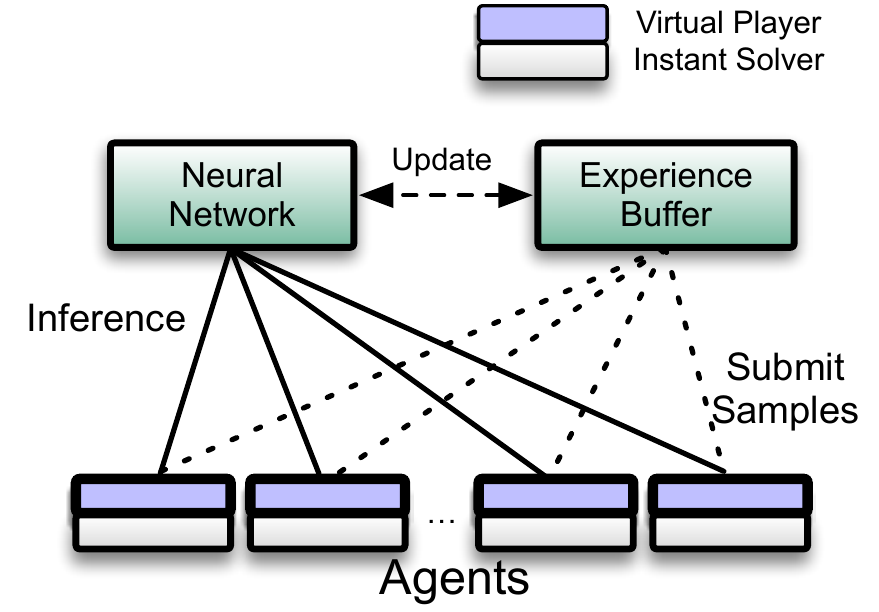}
    \caption{Comyco's Multi-Agent Framework Overview.}
    \label{fig:parallel}
  %   \vspace{-10pt}
\end{figure}

\subsection{Parallel Training}
It's notable that the training process can be designed asynchronously, which is quite suitable for multi-agent parallel training framework. Inspired by the multi-agent training method~\cite{mnih2016asynchronous,huang2018qarc}, we modify Comyco's framework from single-agent training to multi-agent training. 
As illustrated in Figure~\ref{fig:parallel}, Comyco's multi-agent training consists of three parts, a central agent with a NN, an experience replay buffer, and a group of agents with a virtual player and an instant solver. For any ABR state $s$, the agents use virtual player to emulate the ABR process w.r.t current states and actions given by the NN which placed on the central agent, and collect the expert action $\hat{a}$ through the instant solver; they then submit the information containing $\{s, \hat{a}\}$ to the experience replay buffer. The central agent trains the NN by picking the sample batch from the buffer. Note that this can happen asynchronously among all agents. By default, Comyco uses \textbf{12} agents, which is the same number of CPU cores of our PC, to accelerate the training process.

\subsection{Implementation}
We now explain how to implement Comyco. We use TensorFlow~\cite{abadi2016tensorflow} to implement the training workflow and utilizing TFlearn~\cite{tang2016tf} to construct the NN architecture. Besides, we use C++ to implement instant solver and the virtual player. Then we leverage Swig~\cite{beazley1996swig} to compile them as a python class. 
%Comparing the executing time of C++-based instant solver and python-based solver, we find that using C++ will significantly accelerate the training process, with the improvements of 15,000\%. We believe that the solver can be deployed in the community soon. 
Next, we will show more details: Comyco takes the past sequence length $k=8$~(as suggested by \cite{mao2017neural}) and future $7$ video chunk features~(as suggested by \cite{yin2015control}) into the NN. We set learning rate $\alpha=10^{-4}$ and use Adam optimizer~\cite{kingma2014adam} to optimize the model. For more details, please refer to our repository~\footnote{\url{https://github.com/thu-media/Comyco}}.

\section{QoE Metrics \& Video Datasets}
% As mentioned before, recent learning-based ABR work fails to find a proper QoE model for optimizing itselves. Followed by this form of drawbacks, we attempt to investigate an simple yet explainable QoE model for evaluating ABR algorithms. 
% In this work, we start by introducing VMAF, a novel video quality feature which can precisely estimate current perpetual video quality. We then use VMAF metrics to describe the video quality features rather than bitrate that previously models commonly used. Next, we propose AQM~(Attention-based QoE Model), a QoE model that uses self-attention-based DL methods to achieve not only the high performances of estimating QoE metrics but also demystifying the underlying principle of NN's decisions. Finally, we compare the SRCC of AQM with several proposed QoE models under DataBaseIII QoE datasets. Experimental results illustrate that AQM achieves state-of-the-art performance.
Upon constructing the Comyco's NN architecture with considering video content features, we have yet discussed how to train the NN. Indeed, we lack a video quality-aware QoE model and an ABR video dataset with video quality assessment. In this section, we use VMAF to describe the perceptual video quality of our work. We then propose a video quality-aware QoE metric under the guidance of real-world ABR QoE dataset~\cite{duanmu2018sqoe}. Finally, we collect and publish a DASH video dataset with different VMAF assessments.

\begin{figure}
    \centering
    \begin{minipage}{0.32\linewidth}
        \centering
        \subfigure[Video Bitrate: 0.480]{\includegraphics[width=1.0\textwidth]{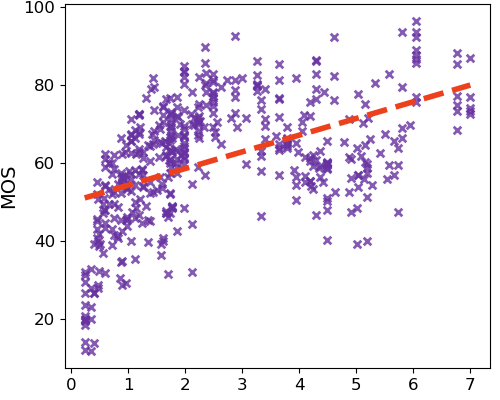}}
    \end{minipage}  
    \begin{minipage}{0.32\linewidth}
        \centering
        \subfigure[SSIM: 0.592]{\includegraphics[width=1.0\textwidth]{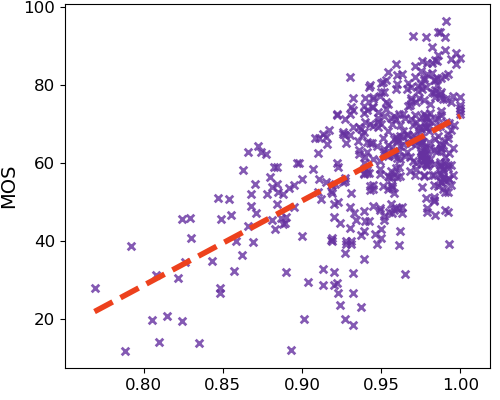}}
    \end{minipage}
    \begin{minipage}{0.32\linewidth}
        \centering
        \subfigure[VMAF: 0.689]{\includegraphics[width=1.0\textwidth]{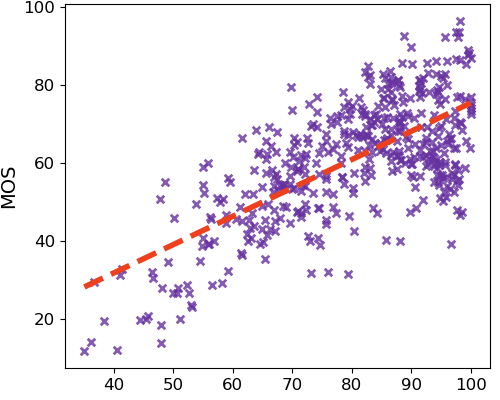}}
    \end{minipage}
    % \begin{minipage}{0.48\linewidth}
    %     \centering
    %     \subfigure{\includegraphics[width=1.0\textwidth]{figs/motivation/vmaf-phone-score}}
    % \end{minipage}     
    % \begin{minipage}{0.24\linewidth}
    %     \centering
    %     \subfigure{\includegraphics[width=1.0\textwidth]{figs/motivation/vmaf-score-fit3}}
    % \end{minipage}
    \vspace{-10pt}
    \caption{Correlation comparison of video presentation quality metrics on the SQoE-III dataset~\cite{duanmu2018sqoe}. Results are summarized by Pearson correlation coefficient~\cite{benesty2009pearson}.}
    \label{fig:bitratetovmaf}
    \vspace{-10pt}
\end{figure}

\subsection{QoE Model Setup}
\label{sec:QoE}
Motivated by the linear-based QoE metric that widely used to evaluate several ABR schemes~\cite{pereira2018cross, yin2015control, akhtar2018oboe, mao2017neural,bentaleb2016sdndash,qin2018abr}, we concluded our QoE metric $\texttt{QoE}_{v}$ as:
\begin{small}
\begin{equation}
\label{eq:general-qoe}
\begin{aligned}
    \texttt{QoE}_{v} &=\alpha\sum_{n=1}^{^{N}}q(R_{n})-\beta\sum_{n=1}^{^{N}}T_{n} \\
                    &+\gamma\sum_{n=1}^{^{N-1}}\left[q(R_{n+1})-q(R_{n})\right]_{+} - \delta\sum_{n=1}^{^{N-1}}\left[q(R_{n+1})-q(R_{n})\right]_{-},
\end{aligned}
\end{equation}
\end{small}
where N is the total number of chunks during the session, $R_n$ represents the each chunk's video bitrate, $T_n$ reflects the rebuffering time for each chunk $n$, $q(R_{n})$ is a function that maps the bitrate $R_n$ to the video quality perceived by the user, $\left[q(R_{n+1})-q(R_{n})\right]_{+}$ denotes positive video bitrate smoothness, meaning switch the video chunk from low bitrate to high bitrate and $\left[q(R_{n+1})-q(R_{n})\right]_{-}$ is negative smoothness. Note that $\alpha$, $\beta$, $\gamma$, $\delta$ are the parameters to describe their aggressiveness. 
%We discuss the choice of $q(R_{n})$ in \S\ref{sec:qrn} and determine the value of each parameter in \S\ref{sec:qoeparam}.

\noindent \textbf{Choice of $q(R_{n})$.}
\label{sec:qrn}
% Estimating QoE via handcrafted features from the client-side has lasted a long history, as several schemes yield a reliable result. Among these features, Video quality assessment~(VQA) plays a crucial part of QoE models. Nevertheless, recent studies pick video bitrate, SSIM or PSNR as the inputs, which fail to characterize the video qualities of the entire video sessions. To prove this point, we test the correlation between MOS and video metrics, including bitrates, SSIMs and different types of VMAF metrics under the DataBaseII QoE datasets. 
To better understand the correlation between video presentation quality and QoE metric, we test the correlation between mean opinion score~(MOS) and video quality assessment~(VQA) metrics, including video bitrate, SSIM~\cite{hore2010image} and Video Multimethod Assessment Fusion~(VMAF)~\cite{rassool2017vmaf}, under the Waterloo Streaming QoE Database III~(SQoE-III)\footnote{SQoE-III is the \emph{largest and most realistic dataset} for dynamic adaptive streaming over HTTP~\cite{duanmu2018sqoe}, which consists of a total of 450 streaming videos created from diverse source content and diverse distortion patterns.}~\cite{duanmu2018sqoe}, where SSIM is a image quality metric which used by D-DASH~\cite{DDASH} and VMAF is an objective full-reference video quality metric which is formulated by Netflix to estimate subjective video quality. Results are collected with Pearson correlation coefficient~\cite{benesty2009pearson} as suggested by~\cite{abar2017machine}. Experimental results~(Fig.~\ref{fig:bitratetovmaf}) show that VMAF achieves the highest correlation among all candidates, with the improvements in the coefficient of 16.39\%-43.54\%. Besides, VMAF are also a popular scheme with great potential on both academia and industry~\cite{aaron2015challenges}. We, therefore, set $q(R_{n})=\texttt{VMAF}(R_{n})$.
%To this end, 

%we believe that using VMAF metric is more capable of characterizing perpetual video quality than previously scheme uses. In this work, 
\begin{small}
\begin{table}[]
\caption{Perfomance Comparison of QoE Models on Waterloo Streaming SQoE-III~\cite{duanmu2018sqoe}}
\begin{tabular}{c|ccc}
\toprule
QoE model  & Type       & VQA & SRCC   \\ \hline
Pensieve's~\cite{mao2017neural} & linear     & - & 0.6256 \\
MPC's~\cite{yin2015control}      & linear     & - & 0.7143 \\Bentaleb's~\cite{bentaleb2016sdndash}  & linear & SSIMplus~\cite{rehman2015display} & 0.6322 \\
Duanmu's~\cite{duanmu2018sqoe}   & linear & - & 0.7743 \\
\rowcolor{mygray}
Comyco's   & linear     & VMAF~\cite{rassool2017vmaf} & \textbf{0.7870} \\ 
% \hline
% P.NATS~\cite{robitza2017modular} & non-linear & O.21, O.22~\cite{robitza2017modular} & 0.8454 \\
% Comyco's DNN   & non-linear & VMAF & \textbf{0.8834} \\
\bottomrule
\end{tabular}
\label{tbl:qoecomaprion}
\vspace{-15pt}
\end{table}
\end{small}

\noindent \textbf{QoE Parameters Setup.}
\label{sec:qoeparam}
Recall that main goal of our paper is to propose a feasible ABR system instead of a convincing QoE metric. 
%Indeed, QoE measurement is another interesting region and 
In this work, we attempt to leverage linear-regression methods to find the proper parameters. Specifically, we randomly divide the SQoE-III database into two parts, 80\% of the database for training and 20\% testing. We follow the idea by~\cite{duanmu2018sqoe} and run the training process for 1,000 times to mitigate any bias caused by the division of data. 
%In general, training time lasts about 500 milliseconds.
As a result, we set $\alpha=0.8469$, $\beta=28.7959$, $\gamma=0.2979$, $\delta=1.0610$. 
%Meanwhile, we also use self-attention NN to investigate 
%0.8469011 * vmaf_sum - 28.79591348 * curr_rebuffer_time + 0.29797156 * \
%                    vmaf_smoothness0 - 1.06099887 * vmaf_smoothness1 - 2.661618558192494
We leverage spearman correlation coefficient~(SRCC), as suggested by~\cite{duanmu2018sqoe}, to evaluate the performance of our QoE model with existing proposed models and the median correlation and its corresponding regression model are demonstrated in Table~\ref{tbl:qoecomaprion}. As shown, $QoE_v$ model outperforms recent work. In conclusion, the proposed QoE model is well enough to evaluate ABR schemes. 
%Because QoE metric
% We also SQoE-III, including video chunk bitrate, SSIM and VMAF-4K under the Waterloo Streaming QoE As illustrated in Fig.~\ref{fig:bitratetovmaf}, we observe that VMAF achieves the highest correlation among all candidates, with the improvements in the coefficient of 16.39\%-43.54\%. Thus, we believe that using VMAF metric is more capable of characterizing the perpetual video quality than previously scheme used. We, therefore, set $q(R_{n})=\texttt{VMAF}(R_{n})$.

\subsection{Video Datasets}
\label{sec:VideoDatasets}
To better improve the Comyco's generalization ability, we propose a video quality DASH dataset involves movies, sports, TV-shows, games, news and MVs. Specifially, we first collect video clips with highest resolution from Youtube, then leverage FFmpeg~\cite{ffmpeg} to encode the video by H.264 codec and MP4Box~\cite{gpac} to \emph{dashify} videos according to the encoding ladder of video sequences~\cite{duanmu2018sqoe,dash}. Each chunk is encoded as 4 seconds. During the trans-coding process, for each video, we measure VMAF, VMAF-4K and VMAF-phone metric with the reference resolution of $1920 \times 1080$ respectively. In general, the dataset contains 86 complete videos, with 394,551 video chunks and 1,578,204 video quality assessments. 
%The dataset will be published later on.

%\input{dataset}
\section{Evaluation}
%In this section, we evaluate Comyco under various network conditions and compare it with previously proposed ABR approaches. 
% Our results answer the following questions:
% \begin{enumerate}[leftmargin=*]
% \item Which scheme is better for Comyco, Comyco with a single output or Comyco with two outputs like traditional buffer-based method?~(\S\ref{sec:bbaoutputs})
% \item How many improvements does dynamic chunk map obtain through Comyco?~(\S\ref{sec:abrbba})
% \item How does Comyco perform? What's the best architecture for Trigger?~(\S\ref{sec:triggereval})
% \item Comparing Comyco with previously proposed approaches on the same network conditions, does Comyco stand for the best method?~(\S\ref{sec:bbapensieve},\S\ref{sec:bbaothers})
% \item Where is the limit of DBB?~(\S\ref{sec:bbalarge})
% \end{enumerate}

\subsection{Methodology}

\label{sec:Implementation}
% \begin{figure}
%     \centering
%     \includegraphics[width=1.0\linewidth]{figs/realworld}
%     \caption{Comyco's Real-world Implementation}
%     \label{fig:deepmpcrw}
% \end{figure}
\textbf{Virtual Player.}~We design a faithful ABR offline virtual player to train Comyco via network traces and video descriptions. The player is written in C++ and Python3.6 and is closely refers to several state-of-the-art open-sourced ABR simulators including Pensieve, Oboe and Sabre~\cite{spiteri2018theory}. 
%We verify the accurateness of the proposed virtual player in Figure~\ref{fig:virtualplayer}. 

\noindent \textbf{Testbed.}~Our work consists of two testbeds. Both server and client run on the 12-core, Intel i7 3.7 GHz CPUs with 32GB RAM running Windows 10. Comyco can be trained efficiently on both GPU and CPU. Detailing the testbed, that includes:
\label{sec:testbed}
\begin{itemize}[leftmargin=*]
    \item[$\triangleright$] \textbf{Trace-driven emulation.}~
    Following the instructions of recent work~\cite{mao2017neural, akhtar2018oboe}, we utilize Mahimahi~\cite{netravali2015mahimahi:} to emulate the network conditions between the client~(ChromeV73) and ABR server~(SimpleHTTPServer by Python2.7) via collected network traces.
    %from our corpus of network traces, along with an 80 ms RTT, 
    
    \item[$\triangleright$] \textbf{Real world Deployment.}~Details are illustrated in \S\ref{real-world}.
\end{itemize}
% \begin{figure}
%     \centering
%     \includegraphics[width=0.4\linewidth]{figs/bitrate}
%     \includegraphics[width=0.4\linewidth]{figs/bufrat}
%     % \begin{minipage}{0.48\linewidth}
%     %     \centering
%     %     \subfigure{\includegraphics[width=1.0\textwidth]{figs/motivation/vmaf-phone-score}}
%     % \end{minipage}     
%     % \begin{minipage}{0.24\linewidth}
%     %     \centering
%     %     \subfigure{\includegraphics[width=1.0\textwidth]{figs/motivation/vmaf-score-fit3}}
%     % \end{minipage}
%     \caption{Comparing the strategy between the virtual player and real \texttt{Dash.js} player. Results show the close correlation of strategy between two players.}
%     \label{fig:virtualplayer}
%     \vspace{-10pt}
% \end{figure}
% \noindent \textbf{Video Datasets.} 
% \label{sec:VideoDatasets}
% In this paper, we use \emph{EnvivioDash3} from the DASH-246 JavaScript reference client~\cite{refclient2016}, the same video dataset commonly used in \cite{mao2017neural, pereira2018cross, akhtar2018oboe}. In details, the video is encoded by the H.264 codec at video bitrates in the range of \{0.3,~0.75,~1.2,~1.85,~2.85,~4.3\} Mbps. The total length of the video is 193 seconds, which is divided into 48 chunks, where each chunk is 4 seconds.

\noindent \textbf{Network Trace Datasets.} 
\label{sec:NetworkDatasets}
We collect about 3,000 network traces, totally 47 hours, from public datasets for training and testing:

\begin{itemize}[leftmargin=*]
\item[$\triangleright$] \textbf{Chunk-level network traces:}~including HSDPA~\cite{riiser2013commute}: a well-known 3G/HSDPA network trace dataset, we use a slide-window to upsampling the traces as mentioned by Pensieve~(1000 traces, 1s granularity); 
%Belgium~\cite{van2016http}: 4G dataset~(40 traces, 1s granularity); 
FCC~\cite{bworld}: a broadband dataset~(1000 traces, 1s granularity); Oboe~\cite{usc-nsl_2018}~(428 traces, 1-5s granularity): a trace dataset collected from wired, WiFi and cellular network connections~(Only for validation.)
%After generating network traces from different types of network traces, we started to convert each saturated trace to a packet-level queue. Inspired by [][], we infer the distribution of packet's sending time as a Poisson process. 
\item[$\triangleright$] \textbf{Synthetic network traces:}~uses a Markovian model where each state represented an average throughput in the aforementioned range\cite{mao2017neural}. We create network traces in over 1000 traces with 1s granularity.

\end{itemize}

% \noindent \textbf{QoE Metrics.}
% \label{sec:QoE}
% In this paper, we use the general QoE metric $QoE_{lin}$~\cite{pereira2018cross, yin2015control, akhtar2018oboe, mao2017neural}, the linear mapping formula which was used by MPC~\cite{yin2015control}, to evaluate Comyco:

% \begin{equation}
% \texttt{QoE}=\sum_{n=1}^{^{N}}q(R_{n})-\mu\sum_{n=1}^{^{N}}T_{n}-\sum_{n=1}^{^{N-1}}\left|q(R_{n+1})-q(R_{n})\right|\label{eq:general-qoe},
% \end{equation}
% where N is the total number of chunks during the session, $R_n$ represents the each chunk's video bitrate, $T_n$ reflects the rebuffering time for each chunk $n$, $q(R_{n})$ is a function that maps the bitrate $R_n$ to the video quality perceived by the user. In this experiment, we set $q(R_{n})=R_{n}$ and $\mu=4.3$.

\noindent \textbf{ABR Baselines.}
\label{sec:abrbaseline}
In this paper, we select several representational ABR algorithms from various type of fundamental principles:
\begin{itemize}[leftmargin=*]
    \item[$\triangleright$] \textbf{Rate-based Approach~(RB)~\cite{jiang2014improving}:}~uses harmonic mean of past five throughput measured as future bandwidth. 
    
    % \item \textbf{Buffer-based Approach~(BB)~\cite{huang2015buffer}:}~dynamically picks next chunk bitrate according to the buffer occupancy.
    
    \item[$\triangleright$] \textbf{BOLA~\cite{spiteri2016bola}:}~turns the ABR problem into a utility maximization problem and solve it by using the Lyapunov function. It's a buffer-based approach. We use BOLA provided by the authors~\cite{spiteri2018theory}.
    
    \item[$\triangleright$] \textbf{Robust MPC~\cite{yin2015control}:}~inputs the buffer occupancy and throughput predictions and then maximizes the QoE by solving an optimization problem. We use C++ to implement \emph{RobustMPC} and leverage $QoE_{v}$~(\S\ref{sec:QoE}) to optimize the strategy.
    %Experimental results on HSDPA network traces show that our RobustMPC implementation improves the average QoE by 4.55\% compared with previous implementation~\cite{hongzimao_2017}.
    
    \item[$\triangleright$] \textbf{Pensieve~\cite{mao2017neural}:}~the state-of-the-art ABR scheme which utilizes Deep Reinforcement Learning~(DRL) to pick bitrate for next video chunks.
    %Pensieve takes the former network status as states and reinforces itself through the interaction with the faithful offline simulator. 
    We use the scheme implemented by the authors~\cite{hongzimao_2017} but retrain the model for our work~(\S\ref{sec:pensieveretrain}).

\end{itemize}

% \begin{figure*}
%     \centering
%     \begin{minipage}{0.49\linewidth}
%         \centering
%         \subfigure{\includegraphics[width=1.0\textwidth]{figs/breakdown/HSDPA}}
%     \end{minipage}  
%     \begin{minipage}{0.49\linewidth}
%         \centering
%         \subfigure{\includegraphics[width=1.0\textwidth]{figs/breakdown/fcc}}
%     \end{minipage}
%     % \begin{minipage}{0.49\linewidth}
%     %     \centering
%     %     \subfigure{\includegraphics[width=1.0\textwidth]{figs/breakdown/oboe}}
%     % \end{minipage}
%     %  \begin{minipage}{0.49\linewidth}
%     %     \centering
%     %     \subfigure{\includegraphics[width=1.0\textwidth]{figs/breakdown/4G}}
%     % \end{minipage}
%     \caption{Comparing AP with existing ABR approaches including BBA and Pensieve on the oboe~\cite{akhtar2018oboe} network traces. Experimental results are shown with the distribution of average bitrate, rebuffering time, average bitrate change for the approaches as well as average QoE.}
%     \label{fig:vmaf}
%   %   \vspace{-10pt}
% \end{figure*}

\begin{figure*}
    \centering
    \begin{minipage}{0.22\linewidth}
        \centering
        \subfigure{\includegraphics[width=1.0\textwidth]{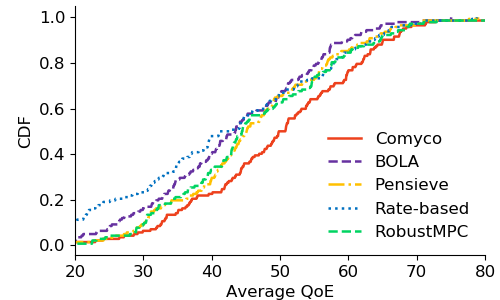}}
    \end{minipage}  
    \begin{minipage}{0.22\linewidth}
        \centering
        \subfigure{\includegraphics[width=1.0\textwidth]{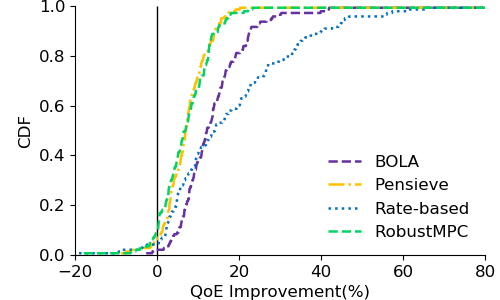}}
    \end{minipage}
    \begin{minipage}{0.46\linewidth}
        \centering
        \subfigure{\includegraphics[width=1.0\textwidth]{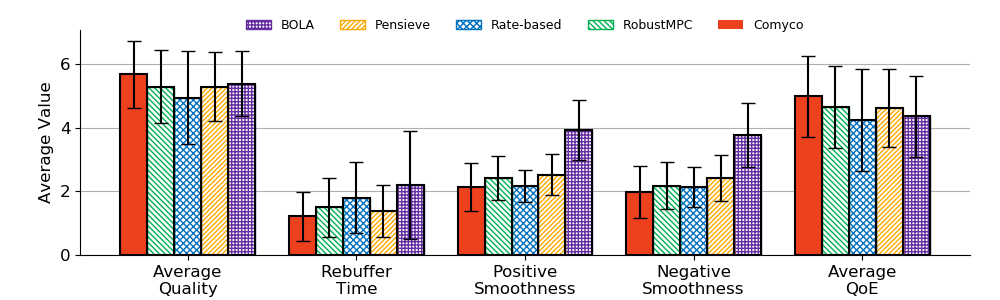}}
    \end{minipage}  
    % \begin{minipage}{0.48\linewidth}
    %     \centering
    %     \subfigure{\includegraphics[width=1.0\textwidth]{figs/oboe/cdf}}
    % \end{minipage}
    %  \begin{minipage}{0.24\linewidth}
    %     \centering
    %     \subfigure{\includegraphics[width=1.0\textwidth]{figs/4G/cdf}}
    % \end{minipage}
    \begin{minipage}{0.22\linewidth}
        \centering
        \subfigure{\includegraphics[width=1.0\textwidth]{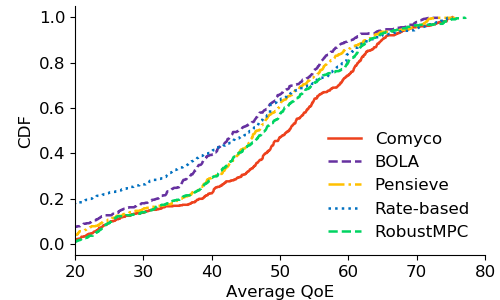}}
    \end{minipage}  
    \begin{minipage}{0.22\linewidth}
        \centering
        \subfigure{\includegraphics[width=1.0\textwidth]{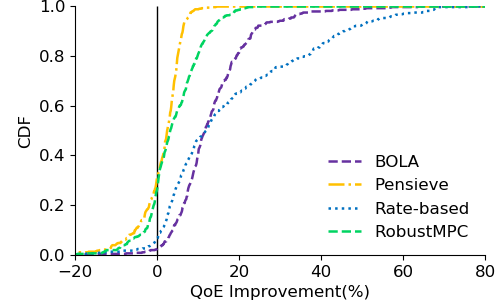}}
    \end{minipage}
    \begin{minipage}{0.46\linewidth}
        \centering
        \subfigure{\includegraphics[width=1.0\textwidth]{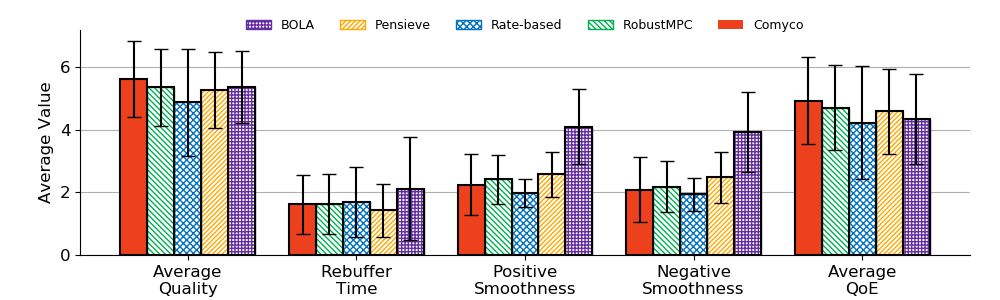}}
    \end{minipage}
    % \begin{minipage}{0.48\linewidth}
    %     \centering
    %     \subfigure{\includegraphics[width=1.0\textwidth]{figs/oboe/diff}}
    % \end{minipage}
    %  \begin{minipage}{0.24\linewidth}
    %     \centering
    %     \subfigure{\includegraphics[width=1.0\textwidth]{figs/4G/diff}}
    % \end{minipage}
    \caption{Comparing Comyco with existing ABR approaches under the HSDPA and FCC network traces. Results are illustrated with CDF distributions, QoE improvement curves and the comparion of several undelying metrics~(\S\ref{sec:QoE}).}
    \label{fig:vmaf}
  %   \vspace{-10pt}
\end{figure*}

\begin{figure}[ht]
    \centering
    \begin{minipage}{0.48\linewidth}
        \centering
        \subfigure[Epochs]{\includegraphics[width=1.0\textwidth]{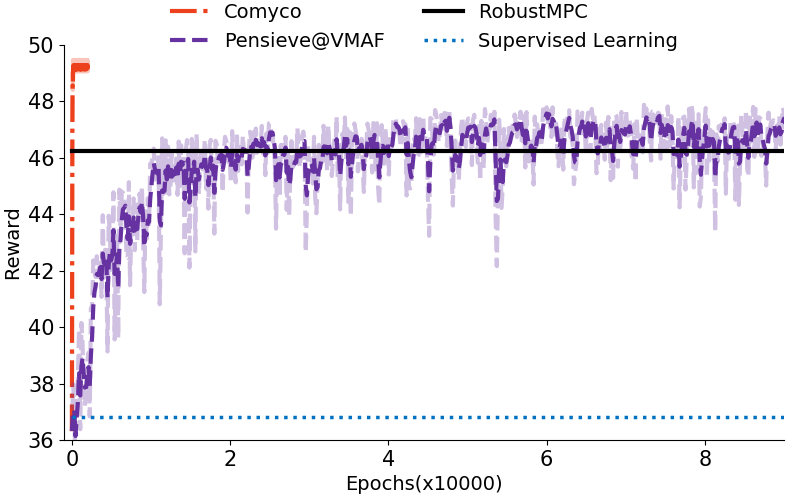}}
    \end{minipage}  
    \begin{minipage}{0.48\linewidth}
        \centering
        \subfigure[Training Time]{\includegraphics[width=1.0\textwidth]{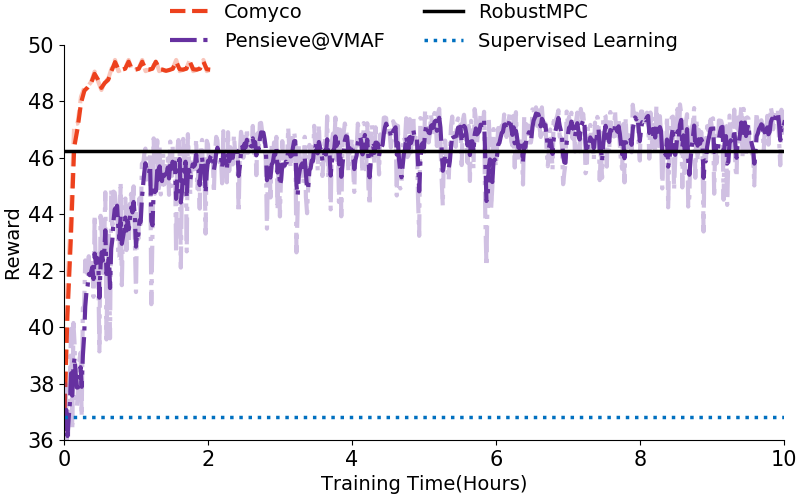}}
    \end{minipage}
    \vspace{-10pt}
    \caption{Comparing the performance of Comyco with Pensieve and Supervised learning-based method under the HSDPA dataset. Comyco is able to achieve the highest performance with significant gains in sample efficiency.}
    \label{fig:comycovspen}
  %   \vspace{-10pt}
\end{figure}

% \subsection{Comyco vs. Pensieve}
% In this part, we setup an offline experiment 

% \textbf{Comyco vs. Pensieve with $QoE_{vmaf}$.}

\subsection{Comyco vs. ABR schemes}
\label{sec:pensieveretrain}
In this part, we attempt to compare the performance of Comyco with the recent ABR schemes under several network traces via the trace-driven virtual player. The details of selected ABR baselines are described in \S\ref{sec:abrbaseline}. We use \emph{EnvivoDash3}, a widely used~\cite{mao2017neural,yin2015control,pereira2018cross,akhtar2018oboe} reference video clip~\cite{dash} and $QoE_v$ to measure the ABR performance.

$\triangleright$ \emph{\textbf{Pensieve Re-training.}}~We retrain Pensieve via our datasets~(\S\ref{sec:NetworkDatasets}), NN architectures~(\S\ref{sec:arch}) and QoE metrics~(\S\ref{sec:QoE}). Followed by recent work~\cite{akhtar2018oboe}, our experiments use different entropy weights in the range of $5.0$ to $1.0$ and dynamically decrease the weight every $1000$ iterations. Training time takes about 8 hours and we show that Pensieve outperforms RobustMPC, with an overall average QoE improvement of 3.5\% across all sessions. Note that same experiments can improve the $QoE_{lin}$~\cite{yin2015control} by 10.5\%. It indicates that $QoE_v$ cannot be easily improved because the metric reflects the real world MOS score. 
%pre-trained Pensieve provided by the authors~\cite{dropboxtraces18}. Recall that we only use Pensieve default trace dataset which provided by the authors~\cite{hongzimao_2017} to train Comyco. So we believe that the trace distribution of ours and Pensieve uses almost identical. As a result, we don't need to retrain Pensieve for our work.

\noindent \textbf{Comparison of Learning-based ABR schemes.}~Figure~\ref{fig:comycovspen} illustrates the average QoE of learning-based ABR schemes on HSDPA datasets. We validate the performance of two schemes respectively during the training. Results are shown with two perspectives including Epoch-Average QoE and Training time-Average QoE and we see about \textbf{1700x} improvement in terms of the number of samples required and about \textbf{16x} improvement in terms of training time required. As expected~(\S\ref{sec:confrontation}), we observe that supervised learning-based method fails to find a strategy, which thereby leads to the poor performance.

 \noindent \textbf{Comyco vs. Existing ABRs.}~Figure~\ref{fig:vmaf} shows the comparison of QoE metrics for existing ABR schemes~(\S\ref{sec:abrbaseline}). Comyco outperforms recent ABRs, with the improvements on average QoE of 7.5\% - 17.99\% across the HSDPA dataset and 4.85\%-16.79\% across the FCC dataset. Especially, Besides, we also show the CDF of the percentage of improvent in QoE for Comyco over existing schemes. Comyco surpasses state-of-the-art ABR approach Pensieve for 91\% of the sessions across the HSDPA dataset and 78\% of the sessions across the FCC dataset. What's more, we also report the performance of underlying metrics including average video quality~(VMAF), rebuffering time, positive and negative smoothness, as well as QoE. We find that Comyco is well behaved on the average quality metric, which improves 6.84\%-15.64\% compared with other ABRs. Moreover, Comyco is able to avoid rebuffering and bitrate changes, which performs as same as state-of-art schemes. 

\begin{figure}
    \centering
    \includegraphics[width=1.0\linewidth]{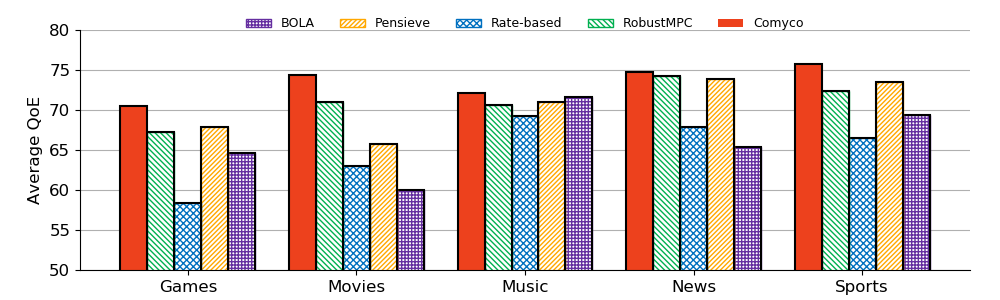}
    \caption{Comparing Comyco with existing ABR approaches under the Oboe network traces and various types of videos.}
    \label{fig:multiple}
    \vspace{-10pt}
\end{figure}

\subsection{Comyco with Multiple Videos}
To better understand how does Comyco perform on various videos, we randomly pick videos from different video types and utilize Oboe network traces to evaluate the $QoE_v$ performances of the proposed methods. Oboe network traces have diversity network conditions, which brings more challenges for us to improve the performance. Figure~\ref{fig:multiple} illustrates the comparison of QoE metrics for state-of-the-art ABR schemes under various video types. We find that Comyco generalizes well under all considered video scenarios, with the improvements on average QoE of 2.7\%-23.3\% compared with model-based ABR schemes and 2.8\%-13.85\% compared with Pensieve. Specifically, Comyco can provide high quality ABR services under movies, news, and sports, which are all the scenarios with frequent scene switches. We also find that Comyco fails to demonstrate overwhelming performance in serving music videos. It's really an interesting topic and we'll discuss it in future work.

\subsection{Ablation Study}
\label{sec:cmcstudy}
In this section, we set up several experiments that aim to provide a thorough understanding of Comyco, including its hyperparameters and overhead. Note that, we have computed the offline-optimal results via dynamic programming and complete network status~\cite{mao2017neural} before the experiment and treated it as a baseline.
%As mentioned before~(\S\ref{sec:arch}), Comyco depends on several hyperparameters, such as future step $N$, entropy weight $\alpha$ and the effective of using experience replay buffer.

% \begin{table}
% \begin{tabular}{c|c|c|c|c}
% \hline
% N & 6     & 7     & 8      & 9              \\ \hline
% Replay & 0.893 & 0.917 & 0.932  & 0.942          \\
% Replay & 0.921 & 0.937 & 0.946  & \textbf{0.960} \\ \hline
% Time(ms) & 8  & 58 & 389 & 2604        \\ \hline
% \end{tabular}
% \end{table}
\newcolumntype{g}{>{\columncolor{mygray}}c}
\begin{small}
\begin{table}
\caption{Comyco with different $N$ and replay strategies.}
\begin{tabular}{c|cccgc}
\toprule
$\alpha=0.001$/N    & 5     & 6     & 7     & 8      & 9              \\ \hline
Replay Off             & 0.883 & 0.893 & 0.917 & 0.932  & 0.942          \\
\rowcolor{mygray}
Replay On              & 0.911 & 0.921 & 0.937 & 0.946  & \textbf{0.960} \\ \hline
TimeSpan(Opt. Off)(ms) & 1.56  & 8.74  & 58.44 & 389.68 & 2604.46        \\ \bottomrule
\end{tabular}
\label{tbl:cmcdeep}
\end{table}
\begin{table}[]
\caption{Comyco with different $\alpha$.}
\begin{tabular}{c|ccgcc}
\toprule
$\alpha$ & 0.1   & 0.01  & 0.001          & 0.0001 & 0     \\ \hline
k=4      & 0.883 & 0.895 & \textbf{0.904} & 0.881  & 0.867 \\ \bottomrule
\end{tabular}
\label{tbl:cmcalpha}
\end{table}
\end{small}

\noindent \textbf{Comparison of different future step N.}~We report normalized QoE and raw time span of Comyco with different N and replay experience strategy in Table~\ref{tbl:cmcdeep}. Results are collected under the Oboe dataset. As shown, we find that experience replay can help Comyco learn better. Despite the outstanding performance of Comyco with N=9, this scheme lacks the algorithmic efficiency and can hardly be deployed in practice. Thus, we choose k=8 for harmonizing the performance and the cost.

\noindent \textbf{Comyco with different $\alpha$.}~Further, we compare the normalized QoE of Comyco with different $\alpha$ under the Oboe dataset. As listed in Table~\ref{tbl:cmcalpha}, we confirm that $\alpha=0.001$ represents the best parameters for our work. Meanwhile, results also prove the effective of utilizing entropy loss~(\S\ref{ses:loss}).

\noindent \textbf{Comyco Overhead.}~We calculate~\cite{molchanov2016pruning} the number of floating-point operations (FLOPs) of Comyco and find that Comyco has the computation of 229 Kflops, which is only 0.15\% of the light-weighted neural network ShuffleNet V2~\cite{ma2018shufflenet} (146 Mflops). In short, we believe that Comyco can be successfully deployed on the PC and laptop, or even, on the mobile.
\begin{figure}
    \centering
    \begin{minipage}{0.49\linewidth}
        \centering
        \subfigure{\includegraphics[width=1.0\textwidth]{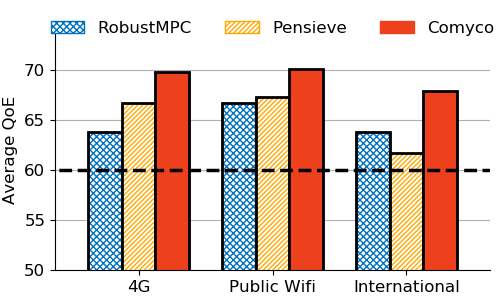}}
    \end{minipage}  
    \begin{minipage}{0.48\linewidth}
        \begin{small}
        \begin{tabular}{c|ccc}
        \toprule
        Network & \begin{tabular}[c]{@{}c@{}}RTT\\ (ms)\end{tabular} & \begin{tabular}[c]{@{}c@{}}$\mu$\\ (KB/s)\end{tabular}  & $\sigma$ \\ \hline
        4G & 65.91   & 325.23  & 53.72     \\ 
        WiFi & 15.58   & 292.98  & 27.65    \\ 
        Inter. & 193.3 & 420.15 & 266.9     \\ \bottomrule
        \end{tabular}
        \label{tbl:nso}
        \end{small}
    \end{minipage}
    \vspace{-10pt}
    \caption{Comparing Comyco with Pensieve and RobustMPC under the real-world network conditions. We take $QoE=60$ as baselines.}
    \label{fig:multiple2}
  %   \vspace{-10pt}
\end{figure}

\subsection{Comyco in the Real World}
\label{real-world}
We establish a full-system implementation to evaluate Comyco in the wild. The system mainly consists of a video player, an ABR server and an HTTP content server. On the server-side, we deploy an HTTP video content Server. On the client-side, we modify \texttt{Dash.js}~\cite{dash} to implement our video player client and we use Chrome to watch the video. Moreover, we implement Comyco as a service on the ABR server. We evaluate the performance of proposed schemes under various network conditions including ~4G/LTE network, WiFi network and international link~(from Singapore to Beijing). Figure~\ref{fig:multiple2} illustrates network status, where $\mu$ is the average throughput measured and $\sigma$ represents standard deviation from the average. 
For each round, we randomly picks a scheme from candidates and summarize the bitrate selected and rebuffering time for each chunk. Each experiment takes about 2 hours. Figure~\ref{fig:multiple2} shows the average QoE results for each scheme under different network conditions. It's clear that Comyco also outperforms previous state-of-the-art ABR schemes and it improves the average QoE of 4.57\%-9.93\% compared with Pensieve and of 6.43\%-9.46\% compared with RobustMPC.
%\vspace{-10pt}
%\noindent \textbf{Online Fine-tuning.}
\section{Related Work}
\label{sec:related}
%In this section, we recent work of ABR including existing DASH Video Datasets and prior ABR schemes.

%\subsection{QoE Datasets}

%\subsection{DASH Video Datasets}

\textbf{ABR schemes.}
Client-based ABR algorithms~\cite{bentaleb2018survey} are mainly organized into two types: model-based and learning-based. 

\emph{\textbf{Model-based.}}~The development of ABR algorithms begins with the idea of predicting throughput. 
%PANDA~\cite{li2014probe} predicts the future throughput for eliminating the ON-OFF steady issue. 
FESTIVE~\cite{jiang2014improving} estimates future throughput via the harmonic mean of the throughput measured for the past chunk downloads. 
%However, due to the lack of throughput estimation method currently, these approaches still result in poor ABR performance. 
%Most video client leverages a playback buffer to store the video content downloaded from #the server temporarily. Thus, 
Meanwhile, many approaches are designed to select the appropriate high bitrate next video chunk and avoid rebuffering events based on playback buffer size observed. BBA~\cite{huang2015buffer} proposes a linear criterion threshold to control the available playback buffer size. %BOLA~\cite{spiteri2016bola} turns the ABR problem into a utility maximization problem and solve it by using the Lyapunov function. 
%However, the buffer-based approach fails to tackle the long-term bandwidth fluctuation problem.
Mixed approaches, e.g., MPC~\cite{yin2015control}, select bitrate for the next chunk by adjusting its throughput discount factor based on past prediction errors and estimating its playback buffer size. 
% Nevertheless, these approaches require careful tuning because they rely on parameters that are quite sensitive to network conditions, resulting in poor performance in unexpected network environments. 
What's more, Akhtar et al.~\cite{akhtar2018oboe} propose an auto-tuning method to improve the model-based ABR's performance.

\emph{\textbf{Learning-based:}}~Several attempts have been made to optimize the ABR algorithm based on RL method due to the difficulty of tuning mixed approaches for handling different network conditions. Pensieve~\cite{mao2017neural} is a system that uses DRL to select bitrate for future video chunks. D-DASH~\cite{DDASH} uses Deep Q-learning method to perform a comprehensive evaluation based on state-of-the-art algorithms. Tiyuntsong optimizes itself towards a rule or a specific reward via the competition with two agents under the same network condition~\cite{huang2018tiyuntsong}.

\noindent \textbf{Imitation Learning meets Networking.}
Imitation learning~\cite{hussein2017imitation} has been widely used in the various fields including networking. Tang et al.~\cite{tanglaoshi} propose real-time deep learning based intelligent network traffic control method to represent the considered Wireless Mesh Network (WMN) backbone via imitation learning. Indigo~\cite{yan2018pantheon} uses DAgger~\cite{ross2011reduction} to train a congestion-control NN scheme in the offline network emulator.

\vspace{-5pt}
\section{Conclusion}
In this work, we propose Comyco, a learning-based ABR system which aim to thoroughly improve the performance of learning-based algorithm. To overcome the sample inefficiency problem, we leverage imitation learning method to \emph{guide} the algorithm to explore and exploit the \emph{better} policy rather than stochastic sampling. Moreover, we construct the video quality-based ABR system, including its NN architectures, datasets and QoE metrics. With trace-driven emulation and real-world deployment, we show that Comyco significantly improves the performance and effectively accelerates the training process. 

\noindent \textbf{Acknowledgement.}~We thank the anonymous reviewer for the valuable feedback. Special thanks to Huang's wife Yuyan Chen, also namely Comyco, for her great support and, happy Chinese valentine's day. This work was supported by the National Key R\&D Program of China (No. 2018YFB1003703), NSFC under Grant 61521002, Beijing Key Lab of Networked Multimedia, and Kuaishou-Tsinghua Joint Project (No. 20192000456).

\newpage

\bibliographystyle{ACM-Reference-Format}
\bibliography{ref}

\end{document}